\documentclass[review]{elsarticle}

\usepackage{lineno,hyperref}
\usepackage{amssymb} %DW
\modulolinenumbers[5]

\journal{Journal of Magnetism and Magnetic Materials}

\begin{document} 

\begin{frontmatter}

\title{Antiferromagnetic CuMnAs: Ab initio description of finite temperature magnetism and resistivity}

\author[MFF,IPM]{David Wagenknecht\fnref{mycorrespondingauthor}}
\fntext[mycorrespondingauthor]{Corresp. author, \textit{Email address:} david@wagenknecht.email, \textit{URL:} wagenknecht.zone}

\author[FZU]{Karel V\'yborn\'y}
\author[MFF]{Karel Carva}
\author[IPM]{Ilja Turek}

\address[MFF]{Department of Condensed Matter Physics, Charles University, Ke Karlovu 5, CZ-12116 Praha, Czech Republic}
\address[IPM]{Institute of Physics of Materials, The Czech Academy of Sciences, \v{Z}i\v{z}kova 22, CZ-61662 Brno, Czech Republic}
\address[FZU]{Institute of Physics, The Czech Academy of Sciences; Cukrovarnick\'a 10, CZ-16200 Praha, Czech Republic}

\begin{abstract}
Noncollinear magnetic moments in antiferromagnets (AFM) lead to a complex behavior of electrical transport, even to a decreasing resistivity due to an increasing temperature.
Proper treatment of such phenomena is required for understanding AFM systems at finite temperatures; however first-principles description of these effects is complicated.
With \textit{ab initio} techniques, we investigate three models of spin fluctuations (magnons) influencing  the transport in AFM CuMnAs; the models are numerically feasible and easily implementable to other studies.
We numerically justified a fully relativistic collinear disordered local moment approach and we present its uncompensated generalization.
A saturation or a decrease of resistivity caused by magnons, phonons, and their combination (above approx.\ 400~K) was observed and explained by changes in electronic structure.
Within the coherent potential approximation, our finite-temperature approaches may be applied also to systems with impurities, which are found to have a large impact not only on residual resistivity, but also on canting of magnetic moments from the AFM to the ferromagnetic (FM) state.
\end{abstract}


\end{frontmatter}

% \linenumbers

\section{Introduction}

The rise of antiferromagnetic spintronics~\cite{Baltz2018} brought about renewed interest in 'old' materials where aspects previously disregarded have now become important.
One of the those materials, which has been known~\cite{Bronger1986} to be an antiferromagnet (AFM) since 1980's and has only recently been identified as a system with locally broken inversion symmetry~\cite{Wadley2013} --- a precondition for the observation of staggered spin-orbit torques~\cite{Zelezny2014}, a novel means of manipulation of magnetic moments in AFMs --- is CuMnAs in its tetragonal phase. 
Detailed understanding of its transport properties is desirable and, with the prospect of using it under conditions of industrial applications~\cite{Jungwirth2016, Wadley2015} allowed by its relatively high N\'eel temperature $T_N\approx 490$~K~\cite{Wadley2015,Uhlirova2019}, effects of chemical and temperature-induced disorder (phonons and magnons) should be included in the model.

The alloy analogy model (AAM) has recently been implemented \cite{DW2017-IEEE, DW2017-SPIE, Drachal2018PRB-DW} within the tight-binding linear muffin-tin orbital (TB-LMTO) method with the coherent potential approximation (CPA) and used to describe FM half-Heusler NiMnSb at finite temperatures \cite{DW2019-JMMM, DW2019-PRB} and also anisotropy of hexagonal systems \cite{DW2019-hexagonal}.
Previous studies of other groups employing the AAM are based on i) the CPA and the Korringa-Kohn-Rostoker (KKR) method with the Kubo-Bastin equation \cite{Ebert, Ebert2011, Kodderitzsch_AHE} and on ii) supercells with the TB-LMTO method and the Landauer-B\"uttiker formula \cite{Glasbrenner, Starikov2018}, but they focus on transition metals.
The disordered local moment (DLM) model within the CPA was used to investigate spin disorder in NiMnSb \cite{Belashchenko2015} and relativistic generalization of the DLM approach within the KKR-CPA-AAM framework was introduced in a study of temperature dependence of magnetic anisotropy~\cite{Staunton2004, Staunton2006}.

On the experimental side, we note that apart from the phase studied in this work, an orthorhombic phase has often been investigated, see Ref.~\cite{Uhlirova2019} for a phase diagram.
This study is focused on tetragonal CuMnAs, which is stabilised by growth on suitably chosen substrates, and we aim on finding a simple yet accurate model of its magnetic disorder to describe finite-temperature electrical transport.
Recent \textit{ab initio} research on tetragonal CuMnAs~\cite{Maca2019} has dealt with transport properties in less complex situations, from the point of view of disorder.
Here, we i) investigate magnetic moments canted towards the FM state by external magnetic field or other techniques~\cite{Zelezny2017} (see Sec.\ \ref{sec_transport_canted}), ii) compare various finite temperature contributions to electrical transport properties, which may play a role in measurements (see Sec.\ \ref{sec_framework} for description of lattice vibrations and Sec.\ \ref{sec_mag_models} for spin fluctuations), and iii) show combined effect of phonons and magnons occuring under real experimental conditions (Sec.\ \ref{sec_transport_impurities}).
Because of the complexity of AFM magnetic structure, we also discuss three models for spin disorder; we present results also for the tilting model (see Sec.\ \ref{sec_mag_models}).
It should be mentioned, that similar aproaches with angular-dependent distribution of the moments are used for ferromagnets also by other authors.
For details see the Refs.\ \cite{Starikov2018} and \cite{Staunton2006}, where authors have used more complex models with a distribution of tilting angles assigned to each given temperature instead of using only one tilting angle for each temperature.

\section{Formalism, Methods, and Models}\label{sec_formalism}
\subsection{Computational framework and CuMnAs}\label{sec_framework}
The fully relativistic TB-LMTO method with the multicomponent CPA and the atomic sphere approximation \cite{IT-book} is used in this study.
For electrical transport, calculations in a framework of the Kubo linear response theory \cite{IT-relativita} with CPA-vertex corrections \cite{KC-multilayers} and a uniform mesh of at least $8\cdot 10^6$ $k-$points was used; increasing the number to $13\cdot 10^6$ resulted in corrections smaller than one percent of the resistivity value.
LSDA+U approach with nonzero Hubbard $U$ is employed for $d$-orbitals of Mn atoms, similarly to \cite{DW2019-JMMM} implemented within the scalar-relativistic TB-LMTO approach \cite{Shick2004}.

Finite-temperature atomic vibrations are approximated by frozen phonons.
Atomic displacements (root-mean-square displacements $\sqrt{\left<u^2\right>}$, later shown in the units of Bohr radius $a_{\rm{B}}$), were related to temperature using the Debye theory with zero-temperature fluctuations omitted \cite{DW2017-IEEE, DW2017-SPIE, Ebert}.
This is a good approximation unless we focus on extreme temperatures such as those occuring in Earth's core \cite{Drchal2017-DW, Drachal2019JMMM-DW}. 
The \textit{spdf-}basis, necessary for inclusion of atomic displacements, was used for most of our calculations.
This study neglects an influence of the Fermi-Dirac distribution modified by finite temperatures.

Both geometry and lattice constants were taken from literature (structure ``II'' in Ref.\ \cite{Maca2019}) and the same values are used for all compositions and temperatures: lattice parameters of bulk P4/nmm CuMnAs are $a = b = 3.82$~\AA~and $c = 6.318$~\AA.
Components of the resistivity tensor $\rho_{xx}$, $\rho_{yy}$, and $\rho_{zz}$ (shown later) correspond to resistivities along $a$, $b$, and $c$, respectively.
Debye temperature of $\Theta_D = 274~\rm{K}$, measured for an orthorhombic sample \cite{Zhang2017}, was used for the lack of experimental data for tetragonal CuMnAs.

In a separate work \cite{Volny2019-CuMnAs}, various types of chemical disorder are discussed in detail while here, we investigate only prototypical and reasonable~\cite{Uhlirova2019} Cu$_{\mathrm{Mn}}$ defect with Cu impurities on Mn sublattices.
Concentrations of this
impurity are stated per formula unit. 

The tetragonal structure of CuMnAs entails large empty spaces between atoms.
To remove possible errors coming from different overlap of the atomic spheres in the in-plane and out-of-plane directions, we also performed reference calculations with empty spheres placed at positions of $[0, 0, 0.5]$ and $[0.5, 0.5, 0.5]$ (with respect to the $a$, $b$, and $c$ lattice directions).
The Wigner-Seitz radius of these empty spheres was set to be smaller by 20~\% compared to other atomic spheres.
Employing eight sublattices (instead of six) increases computational expense; therefore, if not stated otherwise, presented results are obtained without the empty spheres.

Magnetic moments on the two Mn sublattices lie in the $a-b$ plane pointing in two opposite directions with respect to each other \cite{Maca2019}.
In a framework of non-collinear magnetism, we assume two modifications of the AFM ground state:
(a) Magnetic moments may be canted towards each other by an angle $\phi$ so that the moments subtend an angle of ${\pi-2\phi}$, see Fig.\ \ref{g_moments} (a).
The AFM and FM states correspond to $\phi=0$ and $\phi=\pi/2$, respectively.
This approximates a rotation of the moments towards a common direction, e.g., under effect of external magnetic field.
(b) Finite-temperature spin fluctuations are simulated by three models, described in detail in the next subsection \ref{sec_mag_models}. 
We note that the canting and fluctuations may be combined in order to obtain a state influenced by both nonzero magnetic field and finite temperature (not shown here).

\subsection{Models of magnetic disorder}\label{sec_mag_models}

\begin{figure}[!htb]
 \centering
\includegraphics[width=\textwidth]{{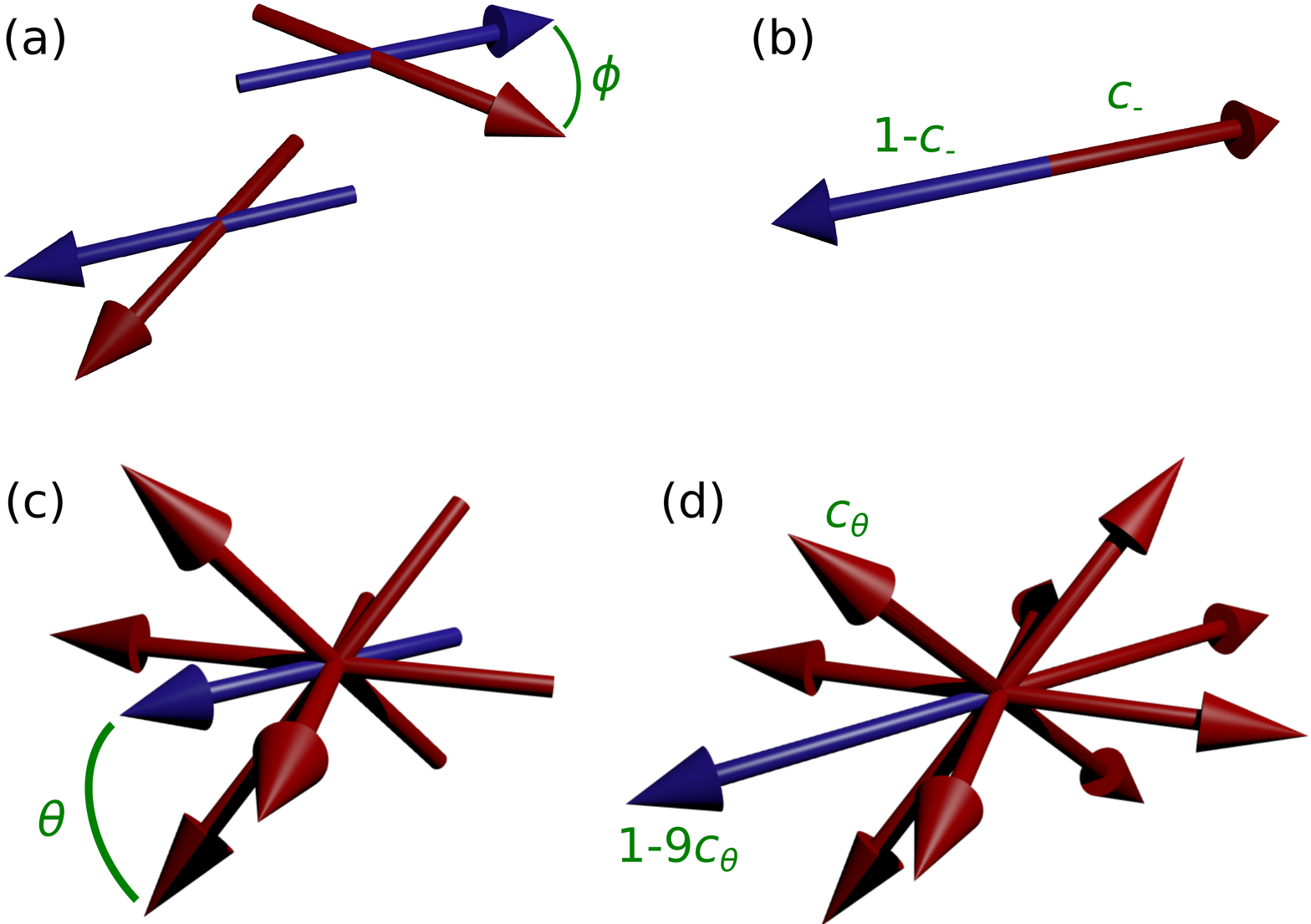}}
 \caption{
 (a) Magnetic moments on two different Mn sublattices are tilted from original AFM (blue) direction by angle $\phi$. Three models of magnetic disorder with original direction of the moments shown by blue arrows:  (b) collinear uDLM, (c) tilting, and (d) tilting uDLM.
 For clarity, (b) -- (d) show only one of the Mn sublattices.
 }
 \label{g_moments}
\end{figure}

There exist rather reliable ways to determine $M(T)$ for a given material, either experimentally or theoretically.
However, to study transport we need to know the distribution of individual spins contributing to total $M$, since it is their variation in space that leads to scattering.
One possibility would be to construct an accurate model in terms of a supercell with atomic spin directions provided for example from atomistic spin dynamics.
Spin directions could also be obtained from a mean-field theory \cite{Wysocki2007}.
However, supercell approach has high numerical demands for transport calculations especially when we need to combine it with the presence of phonons.
The spin-disorder resistivity of Fe and Ni by the supercell technique was investigated in Ref.\ \cite{Wysocki2009}.
Furthermore this construction would still have limitations, for example the incorrect treatment of the low temperature behavior in classical spin dynamics, or the limited accuracy of the mean-field model.
Another possibility is to use the CPA-DLM approach to spin disorder resistivity, which has been found to agree well with the supercell calculations for bcc-Fe, fcc-Ni, and Ni$_2$Mn based compounds \cite{DLM_Kudrnovsky}.
Here we adopt this approach, and a limited number of different spin orientations has to be selected for the alloy analogy model to provide a reasonable description of the spin direction variation that captures the most essential properties. 
We thus employ two existing models here, and also develop a third one.
Their comparison leads to identification of behavior, which is independent on choice of the model.
To describe spin fluctuations, we employ a collinear uncompensated DLM (uDLM) approach and a tilting model, which were used for FM NiMnSb \cite{DW2019-JMMM} and \cite{DW2019-PRB}.
Moreover, we introduce their combination, later called ``tilting uDLM''.
For schematic illustration, see Fig.\ \ref{g_moments} (b)--(d).

The collinear uDLM model is an extension of the widely used DLM method \cite{DLM, DLM2, DLM_Drchal} with two concentrations  $c_+ $ and ${c_- = 1-c_+}$ of magnetic atoms on the same sublattice but with opposite magnetic moments.
The tilting model effectively assumes four mathematically distinguishable atoms (treated within the CPA) having their moments placed on a cone (with the vertex angle $\theta$); axis of the cone corresponds the equilibrium direction \cite{DW2019-JMMM}.
Aiming at having the model as simple as possible, it is sufficient to consider four moments.
For the studied tetragonal system, eight of them (verified for selected cases) led to identical results; a lower number could result in questioning of the proper representation of the symmetry and it would make the tilting model too similar to the uDLM one (therefore, differences may not be easily investigatable).

Systems with relativistic effects have to be investigated numerically with an appropriate distribution of magnetic moments even in the maximally disordered (DLM) state \cite{DLM_Kudrnovsky}.
Such state has zero average magnetization on each sublattice, which can be within the tilting model achieved only by the configuration with $\theta=\pi/2$. However, the tilting angle $\theta=\pi/2$ applied to each magnetic atom (e.g., originally along $x$ direction) would result in moments being in the plane perpendicular to the original direction of the moments ($y-z$ plane).
Therefore, disorder in the original direction (along $x$) is suppressed, and we have introduced an artificial anisotropic distribution in the system, while the moment distribution in the maximally disordered system should be isotropic.

Because of that, we introduce the tilting uDLM model: each fluctuating moment is represented (within the CPA) by ten atoms, one heading towards the original direction, e.g., a vector $(1,0,0)$, second one to the opposite, similarly to the collinear uDLM to $(-1,0,0)$, and around each of them other eight (two times four) form two cones defined by moments tilted from the original directions towards body diagonals, i.e., they would point towards $(\pm 1, \pm 1, \pm 1)$ in a cubic system.
Our system is close to a cubic one; therefore, for a simplicity, we use this fixed vertex angle instead of directions pointing exactly along a diagonal of the tetragonal cell.
Among the three models, this one describes anisotropic material behavior in the best way.
The original direction has concentration of $1-9c_\theta$ and nine others $c_\theta$; unlike the tilting model, two opposite cones for each moments are now constructed with fixed vertex angle.
In contrast to the collinear uDLM approach, the fluctuating moments are now distributed to more directions, which may play a role, e.g., for an anisotropic electrical transport.
Moreover, in contrast with the tilting model, $c_\theta$ may be now increased to $0.1$ (the maximal spin disorder) and it does not lead to a possibility of the moments on different atomic sites being aligned.

We note, that the tilting uDLM model is similar to approach used for fully relativistic investigation of Fe within the DLM model with 26 directions of the moment \cite{DLM_Kudrnovsky}, but now with variable concentrations and applied to tetragonal structure.
In the scalar-relativistic case, the compensated collinear DLM method can be justified analytically (see Appendix of Ref.\  \cite{DLM_Kudrnovsky}) but for the fully-relativistic approach with uncompensated concentrations and non-collinear moments, the treatment of multiple magnetic moments has to be done numerically.

Magnetizations of the Mn sublattices as a function of temperature were calculated by Monte Carlo simulations \cite{Maca2019}.
We related its decrease to our CPA-averaged magnetization calculated as functions of parameters of the models.
Similarly to the FM case \cite{DW2019-JMMM}, it led to a link between temperature values (Monte Carlo simulations) and the parameters ($c_-$, $\theta$, and $c_\theta$).
Because the relation is unambiguous, experimentally well-defined temperature can be used, e.g., as an independent variable in figures.

\section{Results}

\subsection{Electronic structure}

\begin{figure}[!htb]
 \centering
\includegraphics[width=\textwidth]{{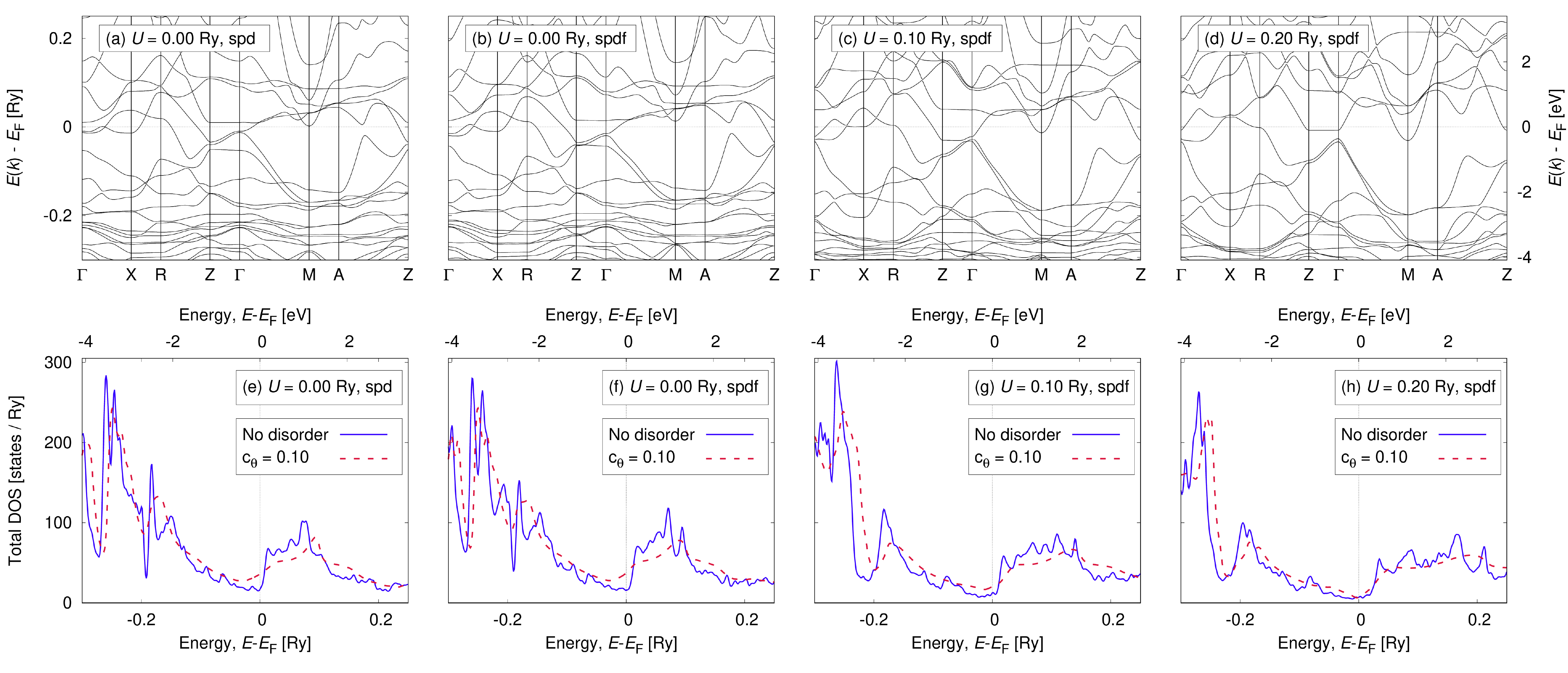}}
 \caption{
 Band structures (a)--(d) and DOS (e)--(h) are calculated for CuMnAs without chemical impurities, atomic vibrations, and empty spheres on interstitial positions.
 DOS for maximal spin disorder within the tilting uDLM approach is shown by red dashed lines; other data are without spin fluctuations.
 See labels for employed basis and Hubbard $U$.
 }
 \label{g_el_structure}
\end{figure}

\begin{figure}[!htb]
 \centering
\includegraphics[width=\textwidth]{{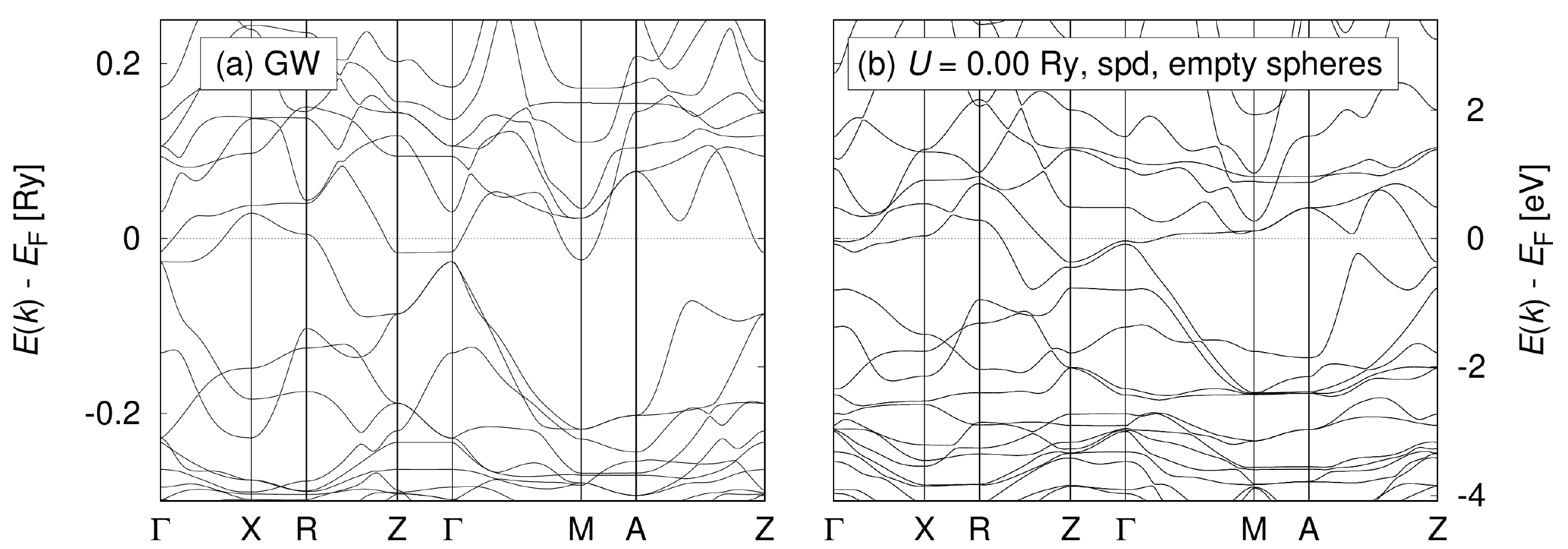}}
 \caption{(a) Band structure calculated in GW approximation. (b) Band structure with empty LMTO spheres at positions of $[0, 0, 0.5]$ and $[0.5, 0.5, 0.5]$ with respect to the \textit{a-b-c} unit cell; other parameters are the same as for Fig.\ \ref{g_el_structure} (a).
 }
 \label{g_el_structure_ES}
\end{figure}

\begin{figure}[h!tpb]
 \centering
\includegraphics[width=0.5\textwidth]{{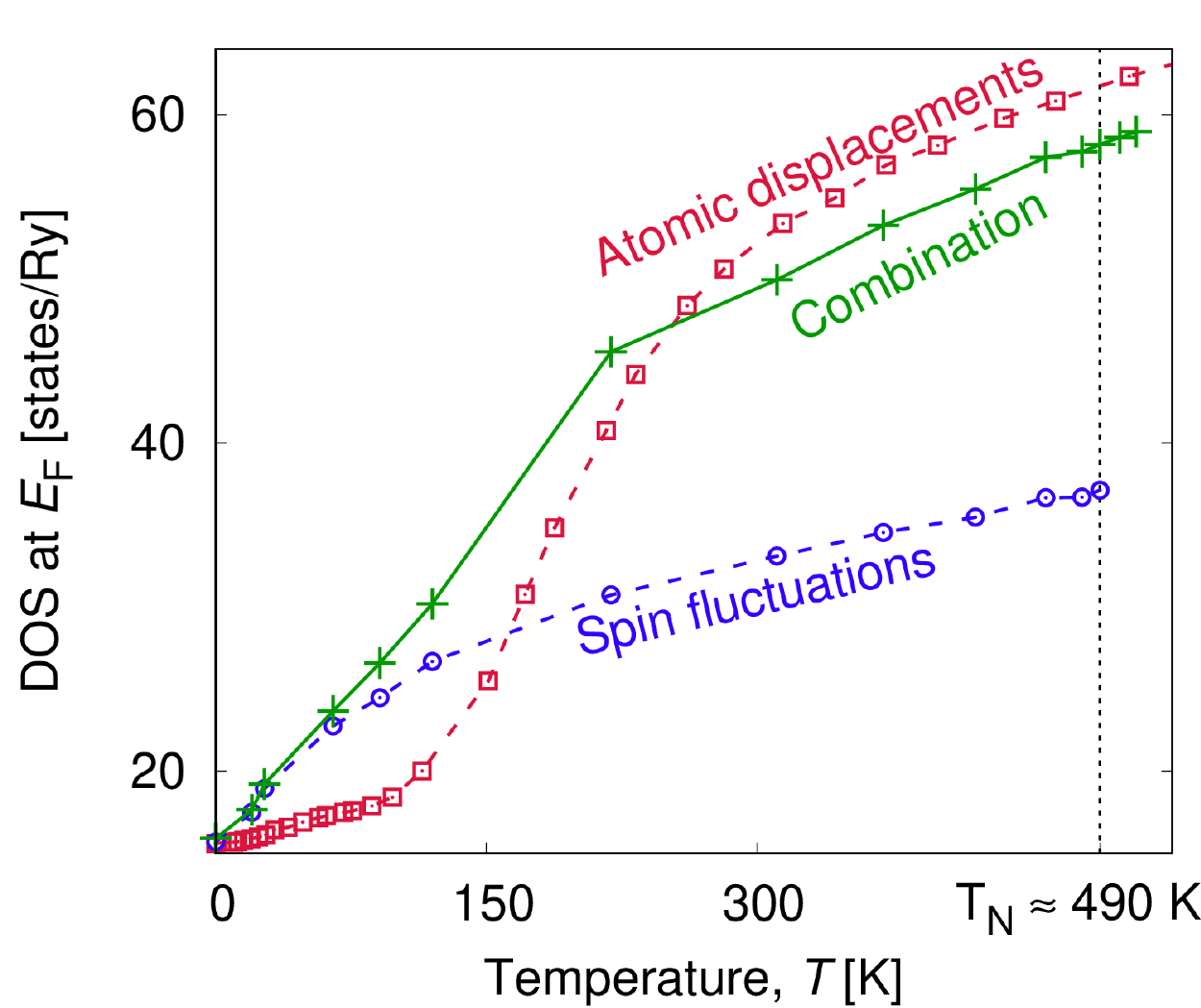}}
 \caption{Increase of DOS at $E_F$ with rising temperature caused by phonons (red squares) and magnons (blue circles) is not additive when compared to their combined effect (green crosses). Spin disorder is treated by the tilting uDLM model.
 }
 \label{g_dos_ef}
\end{figure}

Basis for any further theoretical study is a sound band structure of the perfect crystal.
To this end, there is a broad consensus in literature that tetragonal CuMnAs is a metal with low DOS at the Fermi level (earning it
sometimes the qualifier of a semimetal).
In literature, the most direct experimental probe into its band structure has been photoemission \cite{Veis2018} along with optical spectra obtained by ellipsometry but these integral quantities (as opposed to angular resolved ones) cannot distinguish fine details of the band structure. Such details can, on the other hand, strongly influence transport properties which are sensitive to the situation at the Fermi surface.
In general, the band structures presented below are similar to each other and when related to behavior of real disordered samples, more attention should be paid to other quantities such as electrical transport.

It has been demonstrated \cite{Veis2018} that density functional theory (DFT) calculations give better agreement with optical and photoemission spectra when extended to DFT+$U$ and Fig.\ \ref{g_el_structure} shows how the band structure and DOS depend on the value of $U$ (the two leftmost panels show, at $U=0$, that extending our basis to $spdf$ has only minor effect on the band structure).
Around the Fermi level, the largest differences among these calculations occur close to $M$ and $X$.
From our results, the $spdf$-basis and $U=0.20$~Ry band structure in Fig.\ \ref{g_el_structure}(d) is closest to Ref.\ \cite{Veis2018}.
Also, comparison to nonrelativistic quasiparticle-selfconsistent GW (QSGW) \cite{QuestaalGW} calculation in Fig.\ \ref{g_el_structure_ES}(a) is favourable; the absence of band splitting is caused by the omission of spin-orbit interaction effects.
On the other hand, adding empty spheres (see Fig.\ \ref{g_el_structure_ES} (b) for $U=0.00$~Ry), does not significantly modify the band structure, i.e., it differs only slightly from Fig.\ \ref{g_el_structure} (a).

The DOS at $E_F$ decreases to 61\,\% of its original value when $U$ is changed from 0.00~Ry to 0.20~Ry, see Fig. \ref{g_el_structure} (f)--(h).
In Figs. \ref{g_el_structure} (e)--(h), DOS for maximal spin disorder within the tilting uDLM model ($c_\theta=0.10$) is shown by dashed lines.
Sharp peaks are smeared in presence of spin fluctuations but, unlike to NiMnSb \cite{DW2019-PRB}, DOS at $E_F$ may increase.
This is caused by high DOS above $E_F$ and it may be connected to decrease of $\rho_{zz}$ later shown in Sec.\ \ref{sec_transport_fluctuations}.

With finite temperature effects included, we present a dependence of DOS at $E_F$ on temperature in Fig.\ \ref{g_dos_ef} for $U=0.00$~Ry.
Temperature value (horizontal axis) was obtained separately for atomic vibrations and spin fluctuations (tilting uDLM model).
Based on the large contribution coming from spin fluctuations, this effect is supposed to have large impact on finite-temperature electrical transport and, moreover, it is necessary to properly describe it in the whole relevant temperature range.

\subsection{Magnetic moments and total energy}

Local magnetic moments on each Mn sublattice of undistorted CuMnAs for $U=0.00$~Ry are found to be $3.72\mu_B$ and $3.71\mu_B$ for $spd-$ and $spdf-$calculations, respectively.
This value is increased by nonzero $U$, e.g., it is $4.08\mu_B$ for both bases with $U=0.10$~Ry.
When spin fluctuations are assumed, these values are almost unchanged (up to a few percent) for each direction within the CPA, but the local magnetic moment on each sublattice vanishes monotonically with increasing $\theta$, $c_-$, or $c_\theta$.

\begin{figure}[htbp]
 \centering
\includegraphics[width=\textwidth]{{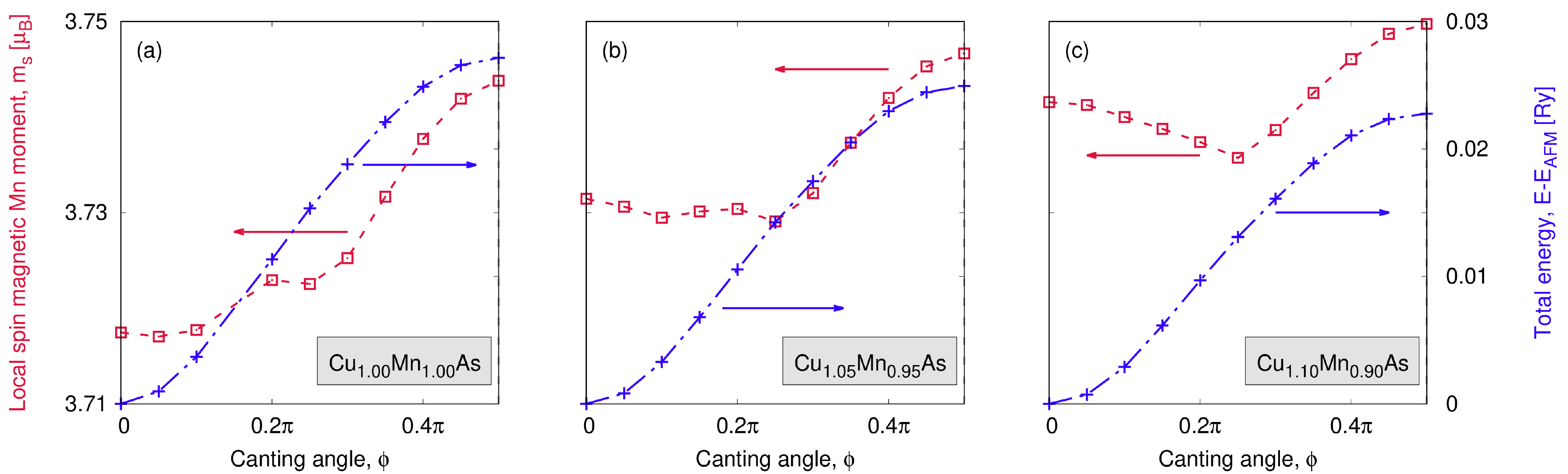}}
 \caption{Mn local moments on the two magnetic sublattices (left axis; red empty squares) and energy difference from the AFM ground state (right axis; blue croses) with tilted magnetic moments present corrects symmetries with respect to $\phi=\pi/2$ and both quantities are influenced only little by increasing Cu$_{\mathrm{Mn}}$ impurity concentration from zero (a) to 5 \% (b), and 10 \% (c).
 Calculated for $U=0.00$~Ry.
 }
 \label{g_magmom}
\end{figure}

Under strong magnetic field, originally antiparallel moments may be forced to cant towards the field direction.
To study this effect, we plot the difference between the total energy of a state with canted moments (in the $a-b$ plane) and the energy of the AFM ground state in Fig.\ \ref{g_magmom}.
Both Mn magnetic moments (Fig.\ \ref{g_magmom} -- left axis) behave equivalently and the energy differences (Fig.\ \ref{g_magmom} -- right axis) were confirmed to have a correct symmetry with respect to $\phi=0.5\pi$.
Mn local magnetic moments are practically unchanged; for stoichiometric CuMnAs, the moments are lowest for the AFM state and there is a minimum in the range of $\phi$ from $0.20\pi$ to $0.3\pi$ for Cu-rich systems.
However, this is not observed for the energy differences, which leads clearly to the AFM configuration being preferred.

Trends visible in Fig.\ \ref{g_magmom} are unchanged regardless whether empty spheres are used or not and also regardless of the basis choice ($spd$ or $spdf$).
The largest change is in the energy difference between the AFM and FM states, e.g., (for stoichiometric undistorted CuMnAs) the $spd-$basis decreases the difference to approx.\ 82~\% and empty spheres to approx.\  44~\% of the original value.

The large difference in the energies implies, that it is difficult to experimentally observe a state with significantly canted moments by applying physically reasonable external fields.
For example, an estimation of the magnetic field required to cant the moments to $\phi=0.2\pi$ results in $B\approx 50$~T.

\subsection{Electrical transport with canted moments}\label{sec_transport_canted}

A large anisotropy of electrical transport is obtained for CuMnAs, regardless orientation of magnetic moments and other conditions.
The longitudinal in-plane resistivity $\rho_{xx}$ may be as many as seven times smaller than the out-of-plane ($\rho_{zz}$, crystallographic direction $c$) one, which is in agreement with Refs.\ \cite{Volny2019-CuMnAs} and \cite{Maca2017}.
Especially $\rho_{zz}$ can reach values of a few hundreds of $\mu\Omega \, \mathrm{cm}$, which is much more than what is usually observed for metallic systems.
These facts are compatible with the layered structure of CuMnAs as well as with its semimetallic nature.

\begin{figure}[htpb]
 \centering
\includegraphics[width=\textwidth]{{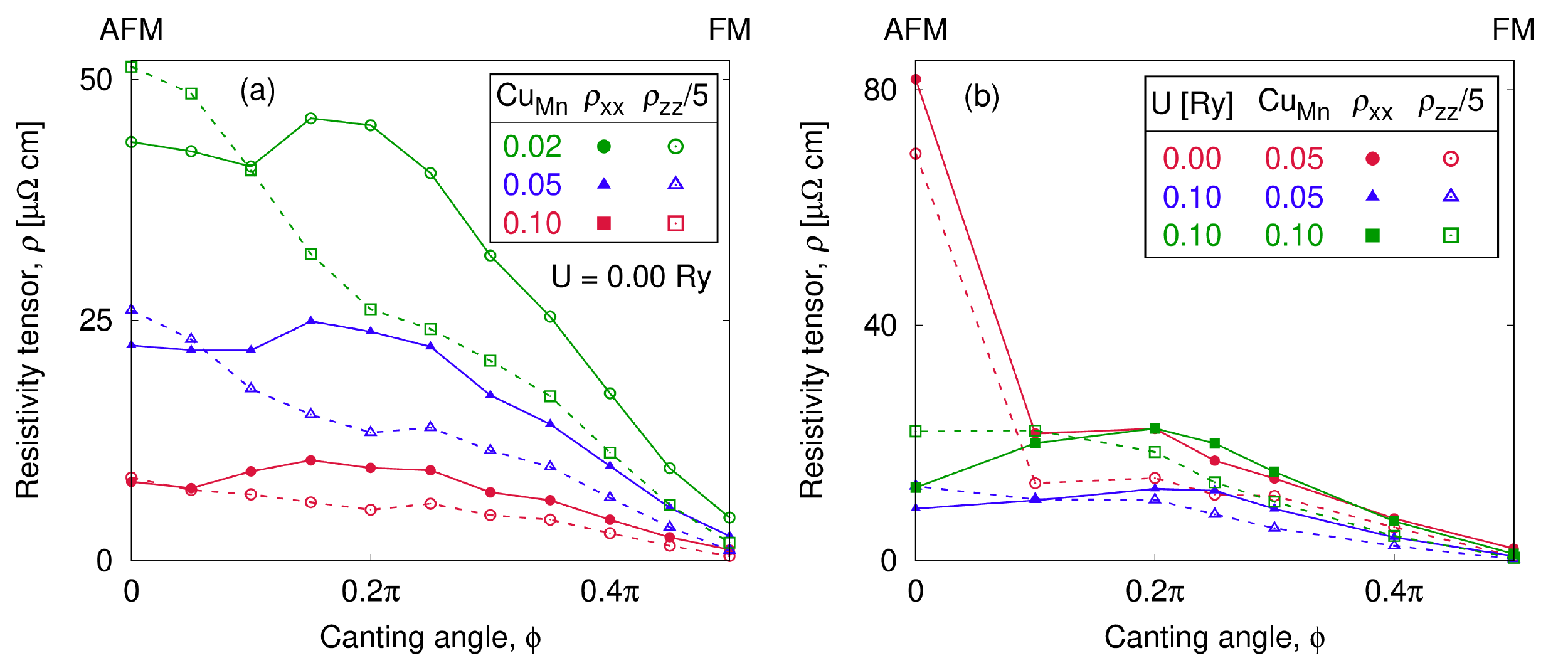}}
 \caption{In-plane $\rho_{xx}$ (full symbols) and out-of-plane $\rho_{zz}$ (empty symbols) are calculated for CuMnAs with canted magnetic moments ($\phi=0$ is original AFM state, $\phi=\pi/2$ gives FM in-plane moments) and the $spd-$basis. Left: Cu atoms on Mn sublattice (Cu$_{\mathrm{Mn}}$) without empty LMTO spheres are assumed for 2~\% (circles), 5~\% (triangles), and 10~\% (squares) of the impurity; $U=0.00$ Ry. Right: For comparison, empty spheres are taken into account for three combinations of $U$ and Cu$_{\mathrm{Mn}}$ concentrations. All $\rho_{zz}$ values are divided by a factor of five.
 }
 \label{g_canting}
\end{figure}

Magnetic moments probably cannot be canted easily by external magnetic fields (see Fig.\ \ref{g_magmom} for energy differences from the AFM state)
and for the manipulation of moments by electric currents \cite{Zelezny2017}, probably, holds the same.
Together with energy analysis, we investigated electrical transport of CuMnAs with canted moments to predict changes, which could be expected.
In Fig.\ \ref{g_canting}, we show resistivities of Cu-rich CuMnAs with magnetic moments canted from the original AFM direction towards the FM orientation.
The canting dramatically reduces both $\rho_{xx}$ and $\rho_{zz}$ and the FM resistivity is much lower than the AFM one.
The in-plane resistivity is nonmonotonic with maxima slightly bellow $\phi=0.25\pi$, which coincides with minima in magnetic moments visible in Fig.\ \ref{g_magmom}.
Fig.\ \ref{g_canting} (a) is obtained without empty spheres, while data in Fig.\ \ref{g_canting} (b) are calculated with the empty spheres on interstitial positions.
The resistivity is increased by alloying but decreased with nonzero $U$.
Increasing $U$ to $0.15$ and $0.20$~Ry reduces resistivities even more (not shown in the Figure).

\subsection{Electrical transport with spin fluctuations}\label{sec_transport_fluctuations}
\begin{figure}[htpb]
 \centering
\includegraphics[width=\textwidth]{{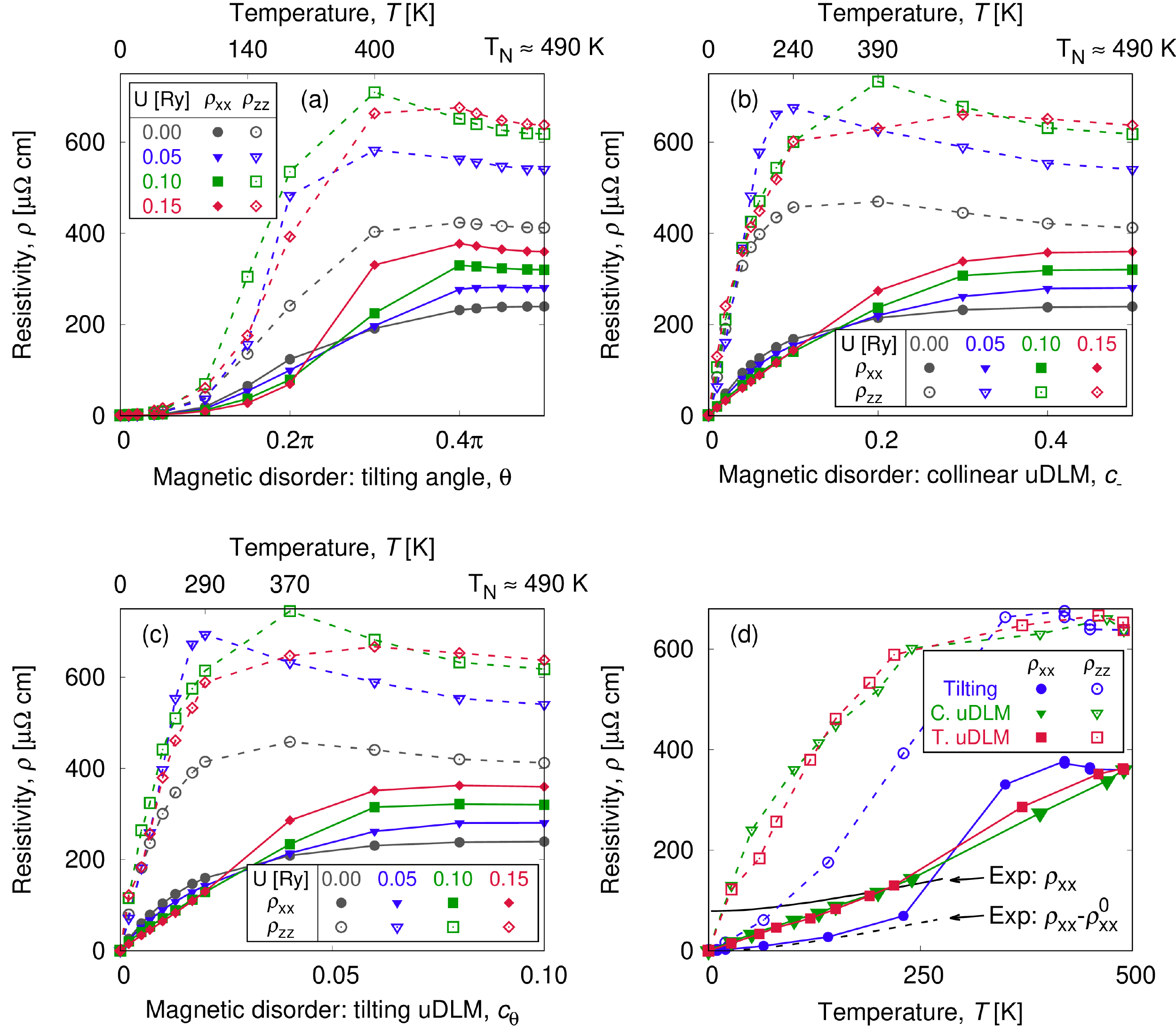}}
 \caption{ Electrical resistivity of CuMnAs without chemical disorder and atomic vibrations is calculated with the $spd-$basis for three models of magnetic disorder: tilting model (a), collinear uDLM model (b), and tilting uDLM model (c). Data for $U$ of 0.00, 0.05, 0.10, and 0.15 Ry are shown by gray circles, blue triangles, green squares, and red diamonds, respectively; $\rho_{xx}$ corresponds to full symbols and $\rho_{zz}$ to empty ones. (d) Results of the three models ($U=0.15$~Ry) plotted as a function of temperature and compared with thin-film measurements \cite{Wadley2013}, see text for details.
 }
 \label{g_rho_mag_dis}
\end{figure}

Electrical resistivity for three models of magnetic disorder is shown in Fig.\ \ref{g_rho_mag_dis}: (a) collinear uDLM, (b) tilting, and (c) tilting uDLM.
The top horizontal axis shows temperature, which corresponds to the decrease of local Mn-sublattice magnetization \cite{Maca2019} for the given parameter of the spin disorder.
The absence of chemical impurities causes $\rho_{xx}=\rho_{zz}=0$ for $\theta = c_- = c_\theta=0$.
Nonzero Hubbard $U$ causes an increase of both $\rho_{xx}$ and $\rho_{zz}$ (except of the tilting model for $\rho_{xx}$ up to room temperature).
Differences between various values of $U$ are small for small spin fluctuations, i.e., for $\theta \lesssim 0.2\pi$, $c_-\lesssim 0.1$, and $c_\theta\lesssim 0.02$; fitting the decrease of Mn-sublattice magnetization \cite{Maca2019} with temperature $T$, these values roughly correspond to $T=$230~K, 240~K, and 220~K, respectively.
Similarity of these temperatures suggests that the three models have similar applicability to real AFM systems.

The collinear uDLM model, Fig.\ \ref{g_rho_mag_dis} (b), assumes the moments only in the in-plane direction, which may give not completely realistic anisotropy of the transport.
Therefore, we calculated maximally disordered DLM state ($c_-=0.5$) with antiparallel moments on each sublattice along $\pm a$, $\pm b$, and $\pm c$ directions: these three cases were found to have elements of the resistivity tensors different by less than 0.1~\% (comparable with numerical errors), which leads to a conclusion, that the collinear uDLM is suitable for study of anisotropy.
Moreover, we present results for the tilting uDLM model in Fig.\ \ref{g_rho_mag_dis} (c): the model has moments more equally distributed (with cubic symmetry instead of the tetragonal one).
Results for the collinear uDLM and tilting uDLM models are almost identical, which numerically justifies the simpler collinear uDLM approach.

Temperature (obtained by fitting of the Mn-sublattice magnetization) is also useful for model comparison, because the resistivity is originally obtained as a function of three different internal parameters ($\theta$, $c_-$, and $c_\theta$).
With temperature increasing from $T=0$, both $\rho_{xx}$ and $\rho_{zz}$ increase linearly up to $T\approx 200$~K, see Fig.\ \ref{g_rho_mag_dis} (d) for $U=0.15$~Ry.

While the collinear and tilting uDLM approaches give almost the same results, they differ from the tilting model, especially at low temperatures.
Fig.\ \ref{g_rho_mag_dis} (d) also compares calculated data with measured thin-film values \cite{Wadley2013} of the planar resistivity (black solid line) and with the residual resistivity of $\rho_{xx}^0=79\,\mu\Omega \, \mathrm{cm}$ subtracted to eliminate an influence of chemical impurities independent on temperature (black dashed line).
Similar values of $\rho_{xx}(T)$ were measured for orthorhombic CuMnAs \cite{Zhang2017}.
$U=0.15$~Ry was chosen for this Figure because of the best correspondence with experiments.
In can be concluded, that the slope of the measured temperature-dependence is well reproduced by the tilting model.
It is similar to behavior of the collinear uDLM model for FM NiMnSb \cite{DW2019-PRB}; however, missing effects of impurities and phonons may be nontrivial, see the next subsection, and because of absence of experimental data, this cannot be studied in greater detail for CuMnAs.
If tetragonal bulk samples of CuMnAs are available, we suggest measuring the out-of-plane resistivity at higher temperatures in order to determine, which model of the spin fluctuations is the most
successful for the studied AFM.
Even if a more advanced model of spin fluctuations were assumed, e.g., a distribution of magnetic moments based on Heisenberg Hamiltonian, it would be difficult to estimate its correctness without comparison with experimental data for conditions of important magnetic disorder.
Therefore, we consider the presented simpler models (neglecting any magnetic short-range order and treated withing the CPA) to be sufficient in the temperature range of available experimental data.

The general trend of decreasing resistivity from room to the N\'eel temperature correlates with and can probably be attributed to large increase of DOS at the Fermi level caused by spin fluctuations, see previous Figs.\ \ref{g_el_structure} and \ref{g_dos_ef}.
Despite the fact that there is no straightforward relation between DOS at the Fermi level and electrical transport, it can be addressed, e.g., by model calculations \cite{Brouers1975}.

\subsection{Electrical transport with phonons and impurities}\label{sec_transport_impurities}
\begin{figure}[htpb]
 \centering
\includegraphics[width=0.7\textwidth]{{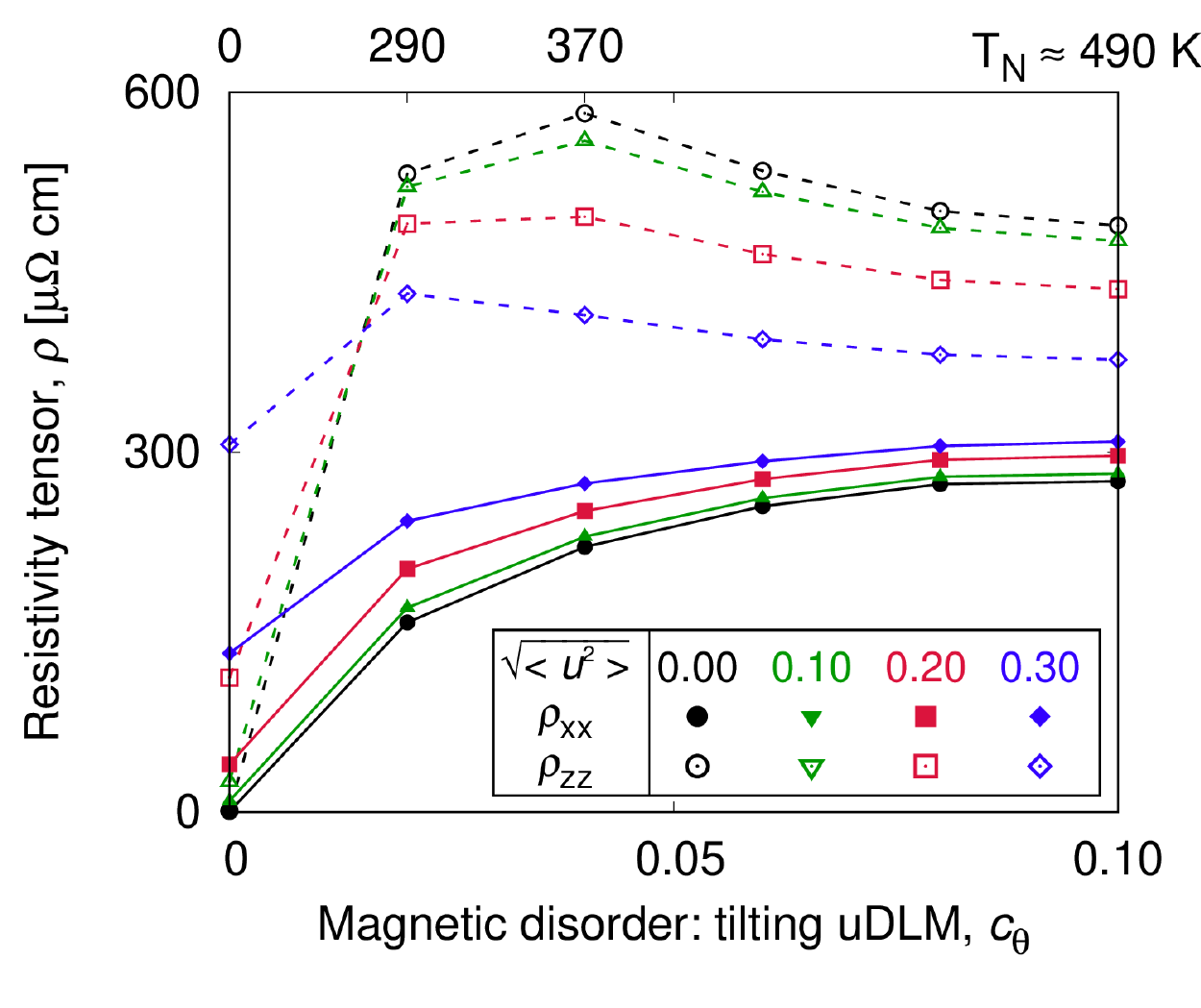}}
 \caption{ 
  Saturation and decrease of resistivities is obtained also for the combined effect of spin fluctuations and atomic displacements (values of $\rho_{zz}$ for $\sqrt{\left<u^2\right>}$ are in $a_{\rm{B}}$, calculated with the $spdf-$basis and $U=0.00$~Ry).
 }
 \label{g_rho_mag_dis_disp}
\end{figure}

To study the decrease of resistivity due to temperature-induced increase of the DOS at $E_F$ in more details, we present a combined effect of spin fluctuations and atomic displacements in Fig.\ \ref{g_rho_mag_dis_disp}.
In contrast to Fig.\ \ref{g_rho_mag_dis}, it was obtained with the $spdf-$basis required because of the displacements; it is the reason for slightly different values for $\sqrt{\left<u^2\right>}=0.00\, a_{\rm{B}}$.
The displacements separately cause a large increase of the DOS at the Fermi level (Fig.\ \ref{g_dos_ef}) and, consequently, saturation of $\rho_{xx}$ and decrease of $\rho_{zz}$ for $\sqrt{\left<u^2\right>} \gtrsim 0.40\, a_{\rm{B}}$ compared to smaller displacements (see also Tab.\ \ref{t_rho_u}).
Similarly to the pure effect of magnons, the combined effect of the spin fluctuations and the displacements also leads to the saturation of $\rho_{xx}$ and to the decrease of $\rho_{zz}$.
Similar behavior was observed also for the tilting model having $\rho_{zz}$ lesser by approx. 15~\% in comparison to Fig.\ \ref{g_rho_mag_dis_disp}.

Nonzero values of $U$ increase resistivities when nonzero atomic vibrations are assumed, while a decrease is observed for CuMnAs with realistic \cite{Uhlirova2019} Cu$_{\mathrm{Mn}}$ impurity concentrations, see Tables \ref{t_rho_u} and \ref{t_rho_c} ($spdf-$calculations).
Both kinds of disorder explain well the large changes in resistivity and its anisotropy, which makes identification of contributions in experimental data difficult.
Ani\-so\-tro\-py $\rho_{zz}/\rho_{xx}$ is increasing with $U$, decreasing with $\sqrt{\left<u^2\right>}$ and it depends strongly especially on chemical composition.
For the atomic displacements (Tab.\ \ref{t_rho_u}), we also observe saturation of $\rho_{xx}$ and decrease of $\rho_{zz}$ for large temperatures, which can be attributed to the increasing DOS at $E_F$, similarly to the case of magnetic disorder.
The in-plane residual (4 K) resistivity in Ref.\ \cite{Wadley2013} was measured about $\rho_{xx}^0 \approx 80\,\mu\Omega \, \mathrm{cm}$ for epitaxial tetragonal CuMnAs indicating even larger chemical disorder than presented in Tab.\ \ref{t_rho_c}.
The same study reported room-temperature resistivity of $160\,\mu\Omega \, \mathrm{cm}$, which would agree with the effect of phonons and magnons, if the impurities are neglected (Fig.\ \ref{g_rho_mag_dis_disp}).
Comparing Fig.\ \ref{g_rho_mag_dis_disp} and Tab.\ \ref{t_rho_u}, combined influence of various scattering mechanisms clearly deviates from Matthiessen's rule; therefore, such room-temperature value may be realistic, but the treatment of finite-temperature effects and chemical impurities together remains a topic for further study \cite{Volny2019-CuMnAs}.

\begin{table}
\caption{ Electrical resistivities of CuMnAs without chemical disorder and spin fluctuations are given in $\mu\Omega \, \mathrm{cm}$  and calculated with the $spdf-$basis.}
\label{t_rho_u}\centering
  \begin{tabular}{|c c||cc|cc|}\hline
    $\sqrt{\left<u^2\right>}$ & T &  \multicolumn{2}{c|}{$U=0.00$ Ry}&	\multicolumn{2}{c|}{$U=0.20$ Ry}\\
     $[a_{\rm{B}}]$ & [K]	&	$\rho_{xx}$ & $\rho_{zz}$  & $\rho_{xx}$ & $\rho_{zz}$  \\ \hline
    0.1 & 50  &	 9&	27&	     11&  50	\\
    0.2 & 110  &	 40&	113&	 51&	206	\\
    0.3 & 185  &	 133&	307&	 142&	455	\\
    0.4 & 290  &	 198&	251&	282&	468	\\
    0.5 & 415  &	 203&	214&	283&	319	\\ 
    0.6 & 570  &	 209&	204&	271&	267	\\ \hline
  \end{tabular}  
\end{table}

\begin{table}
\caption{ Electrical resistivities of Cu-rich CuMnAs without atomic vibrations and spin fluctuations are given in $\mu\Omega \, \mathrm{cm}$ and calculated with the $spdf-$basis.}
\label{t_rho_c}\centering
  \begin{tabular}{|c||cc|cc|}\hline
    &  \multicolumn{2}{c|}{$U=0.00$ Ry}&	\multicolumn{2}{c|}{$U=0.20$ Ry}\\
    Cu$_{\mathrm{Mn}}$	&	$\rho_{xx}$ & $\rho_{zz}$  & $\rho_{xx}$ & $\rho_{zz}$  \\ \hline
    2~\%&	 10&	51&	     2&  6	\\
    5~\%&	 23&	131&	 4&	32	\\
    10~\%&	 40&	317&	15&	285	\\ \hline
  \end{tabular}  
\end{table}

\section{Conclusions}
We have presented an \textit{ab inito} investigation of electronic structure and electrical transport in tetragonal AFM CuMnAs.
For a treatment of nonzero temperatures, we have employed the CPA-AAM with frozen phonons.
Real magnetic disorder is approximated by three models, which are simple and, therefore, easy to implement to other studies, but still reasonably accurate; due to their nature, only results independent on the choice of the model can be considered as physically relevant.
Because electrical resistivity obtained for uDLM and tilting-uDLM models (which differ in directions of the moments) agree with each other and with experimental data, we consider these models to be more appropriate for the AFM CuMnAs than the tilting one.
To obtain more realistic description of the finite temperature behavior, e.g., Monte Carlo simulations and many different directions of the moments could be used.

Both phonons and magnons are treated within the CPA, there is no significant increase of numerical expenses, compared to zero-temperature calculations.
TB-LMTO band structure in LSDA$+U$ approach is compared to the GW results and the best correspondence is found for $U=0.20$ Ry.
Substantial canting of magnetic moments by external magnetic field is probably not achievable; therefore, related decrease of resistivity should not play an important role in experiments.
Saturation of $\rho_{xx}$ and decrease of $\rho_{zz}$ in CuMnAs was observed for temperatures above room temperature, which is not common for metallic systems, but it can be explained by increasing DOS at $E_F$.

For reasonable conditions of room temperature and chemical disorder featured by additional 5~\% of Cu atoms on Mn-sublattices, the largest separate contribution (among impurities, magnons, and phonons) to the resistivity is coming from spin fluctuations.
The tilting model of the spin fluctuations agrees well with the measured slope of $\rho_{xx}(T)$.
We considered also collinear and tilting uDLM approaches; they appear to overestimate the resistivity in the low temperature regime, since already for low temperatures they include moments oriented drastically differently from each other, leading to a strong scattering.
For temperatures $T \gtrsim 0.5 T_N$, the results from all three methods are rather close and we have also numerically justified the collinear uDLM, including the anisotropy of the resistivity.
Therefore, even the most simple collinear uDLM method thus provides a good description for the cases of strong spin disorder.

\section*{Acknowledgements}
Financial support from the Czech Science Foundation is acknowledged by DW, KC, and IT (Grant No.\ 18-07172S) and KV (Grant No.\ 19-28375X).
This work was also supported by The Ministry of Education, Youth and Sports from the Large Infrastructures for Research, Experimental
Development and Innovations project ``IT4Innovations National Supercomputing Center – LM2015070'' and grant No.\ LM2018110 and LNSM-LNSpin.
Access to computing and storage facilities owned by parties and projects contributing to the National Grid Infrastructure MetaCentrum provided under the programme ``Projects of Large Research, Development, and Innovations Infrastructures" (CESNET LM2015042), is greatly appreciated as well as the support of EU FET Open RIA Grant No.\ 766566.

\section*{References}

% \bibliography{references}

\end{document}